# $NaV_2O_4$: a Quasi-1D Metallic Antiferromagnet with Half-Metallic Chains

K. Yamaura,[1,2,*] M. Arai,[1] A. Sato,[1] A.B. Karki,[3] D.P. Young,[3] R. Movshovich,[4] S. Okamoto,[2] D. Mandrus,[2] E. Takayama-Muromachi[1]

[1] National Institute for Materials Science, 1-1 Namiki, Tsukuba, Ibaraki 305-0044, Japan
[2] Materials Science and Technology Division, Oak Ridge National Laboratory, Oak Ridge, TN 37831, USA
[3] Department of Physics and Astronomy, Louisiana State University, Baton Rouge, LA 70803, USA
[4] Los Alamos National Laboratory, Los Alamos, NM 87545, USA

$NaV_2O_4$ crystals were grown under high pressure using a NaCl flux, and the crystals were characterized with X-ray diffraction, electrical resistivity, heat capacity, and magnetization. The structure of $NaV_2O_4$ consists of double chains of edge-sharing $VO_6$ octahedra. The resistivity is highly anisotropic, with the resistivity perpendicular to the chains more than 20 times greater than that parallel to the chains. Magnetically, the intrachain interactions are ferromagnetic and the interchain interactions are antiferromagnetic; 3D antiferromagnetic order is established at 140 K. First principles electronic structure calculations indicate that the chains are half metallic. Interestingly, the case of $NaV_2O_4$ seems to be a quasi-1D analogue of what was found for half-metallic materials.

Quasi-1D metals have intrigued condensed matter physicists for decades, and they still continue to fascinate both due to their propensity toward Fermi surface instabilities such as density waves and to their ability to form Luttinger liquid and other non-Fermi-liquid ground states. Efforts to elucidate quasi-1D phenomena are an important topic of ongoing experimental and theoretical research.

Recent examples of novel quasi-1D behavior in vanadates include $\beta$-Na$_{0.33}$V$_2$O$_5$, which was found undergo a superconducting transition (believed to be non-BCS) by applying pressure [1], and BaVS$_3$, which showed Luttinger liquid behavior in its photoemission spectra [2]. With these examples in mind, we paid particular attention to the $3d^1$ system and related compounds to search for a novel electronic state. Specifically, we chose the electrically insulating compound CaV$_2$O$_4$ comprising double-chains of edge-sharing VO$_6$ octahedra (known as the CaFe$_2$O$_4$ structure [3]) as a subject, and conducted aliovalent substitution of Ca$^{2+}$ by Na$^{1+}$ to increase the population of the $3d^1$ state. Here we report the initial synthesis of NaV$_2$O$_4$ using a high pressure method. X-ray diffraction confirmed that this mixed-valent V$^{3.5+}$ ($3d^{1.5}$) compound is indeed isostructural to CaV$_2$O$_4$. Its quasi-1D metallic nature and an antiferromagnetic (AF) transition ($T_N$ ~140 K) were confirmed by magnetic and transport measurements. Intriguingly, the compound maintains a quasi-1D metallic state down to 40 mK, and the metallicity coexists with magnetic order. This is surprising because in the presence of magnetic interactions a quasi-1D metal is generally unstable against the formation of a spin density wave. Moreover, a small amount of disorder can localize the charge carriers, transforming a one-dimensional metal into an insulator [4]. Even though the electronic system of NaV$_2$O$_4$ is not sufficiently anisotropic enough to warrant the occurrence of quasi-1D instabilities, as far as we know there are very few materials exhibiting metallic nature below a well defined AF transition with an anisotropic crystal structure. The metallic ground state of NaV$_2$O$_4$ is therefore quite surprising. In this Letter, we report basic physical properties of this new quasi-1D metallic vanadium oxide.

Single crystals of NaV$_2$O$_4$ were grown by a NaCl flux technique in a high-pressure furnace that was able to maintain 6 GPa throughout the crystal growth [5]. The crystals grew as shiny black needles or platelets as shown in Fig. 1a. Single crystal diffraction data were obtained on a Bruker SMART APEX (MoK$_\alpha$ $\lambda$=0.71069Å) diffractometer and were analyzed using the SHELXL-97 analysis software [6]. The *R*-factors in the final refinement, 2.24% ($R_p$) and 5.83% ($R_{wp}$), indicate good convergence. The study clearly confirmed the CaFe$_2$O$_4$ structure for the crystal [3]. A brief summary of the structural parameters appears in Table 1, and Figs. 1b and 1c show, respectively, a schematic view of the crystal structure and an illustration of the VO$_6$ double chains that run along the *b*-axis.

Magnetization measurements were performed on the 1.779 mg crystal shown in Fig. 1a using a Quantum Design MPMS between 1.8 K and 350 K and in an applied field of 50 kOe. Electrical resistivity measurements were performed both parallel and perpendicular to *b* from 2 K to 300 K using a Quantum Design PPMS and from 40 mK to 4 K using a $^3$He/$^4$He dilution refrigerator. The crystals were large enough for 4-wire measurements to be used in all cases. 20-30 μm Pt wires were affixed to the samples using silver epoxy. Voltage leads were separated by 0.2-0.5 mm for measurements parallel to *b*, and 0.1-0.2 mm for measurements perpendicular to *b*. Magnetoresistance (MR) measurements were performed in fields to 90 kOe at temperatures of 50 K, 150 K, and 200 K.

Specific heat measurements were performed on a polycrystalline sample prepared from Na$_2$O$_2$ and V$_2$O$_3$ heated at 1500 ºC for 1 h at 6 GPa. The measurement employed a relaxation technique. Measurements were performed between 2 K and 160 K, and with and without an applied magnetic field of 70 kOe. Additional data were collected on a collection of 10 crystals with Ag paint thermal connections between them in a dilution refrigerator between 70 mK and 3 K. Seebeck coefficient measurements were performed in a Quantum Design PPMS using constantan as a reference material. The absolute thermopower of NaV$_2$O$_4$ was obtained by subtracting the thermopower of constantan from the measured data.

Since the vanadium ions have a non-integral valence (V$^{+3.5}$; 3$d^{1.5}$), we carefully examined the crystallography data for each vanadium site (see Table 1) to investigate possible non-uniformity of charge distribution. The two site surroundings, however, show no significant differences. For example, the average distance to the ligand O was 1.97(1) Å and 1.98(2) Å, and the bond valence sum was +3.4 and +3.3 for the V1-site and the V2-site, respectively [7]. Moreover, the distortion factor (defined as the ratio of the longest to the shortest distance) of each of the $VO_6$ octahedron was 1.020 and 1.024 for V1 and V2, respectively, being much lower than the factors of other isostructural materials [8]. It should be stressed here that the $t_{2g}$ electronic system shows a clear AF ordering (shown later), however the X-ray results are not suggestive of expected orbital ordering or crystal-field splitting [9,10]. Low temperature X-ray study might provide further aspects to the issue.

Magnetic susceptibility ($\chi$) of $NaV_2O_4$ is plotted in Fig. 2a, and inverse susceptibility ($1/\chi$) is plotted in Fig. 2b. A clear transition, consistent with antiferromagnetism, is evident near 140 K. To parametrize the data, a Curie-Weiss fit was performed at high temperatures (> 250 K). Fits were performed with and without a temperature-independent paramagnetic term, $\chi_0$. For $\chi_0 = 0$, the best fit yielded 1.99 $\mu_B$/V for the effective moment per vanadium ion, and +118(3) K for the Weiss temperature, $\Theta_W$, suggesting predominantly ferromagnetic (FM) interactions. Interestingly, the fitted moment of 1.99 $\mu_B$/V is close to that expected for a purely localized model, 2.35 $\mu_B$/V [11]. For $\chi_0 \neq 0$, the fit yielded 2.70 $\mu_B$/V and +55(8) K at $\chi_0 = -9.70 \times 10^{-4}$ emu/mole of V. It should be kept in mind, however, that the applicability of the Curie-Weiss law to an itinerant system is not theoretically well-justified, although in practice often yields useful results.

Although the positive $\Theta_W$ from the Curie-Weiss fits suggests FM interactions, but the shape of the transition is clearly antiferromagnetic. Furthermore, isothermal magnetization data (Figs. 2c and 2d) did not show any trace of FM behavior such as a spontaneous magnetization or magnetic hysteresis. We therefore tentatively conclude that strong FM intrachain interactions co-exist with weaker AF interchain interactions, resulting in 3D AF order near 140 K. Further studies using NMR spectroscopy, high-field magnetization, and neutron scattering are underway.

The behavior of the susceptibility below the AF transition indicates that the ordered moments are collinear with the *b*-axis, as the susceptibility with the field applied parallel to *b* drops sharply below the transition. Further checks were made to confirm this conclusion. For example, the $H \perp b$ data for the crystal rotated 90 degrees along the *b*-axis (not shown) were found equivalent to the $H \perp b$ data obtained without the rotation. Moreover, the *M* vs. *H* data in Figs. 2c and 2d show magnetic anisotropy below the ordering temperature: the curves at 150 K and 300 K did not show orientation dependence, while the 10 K curves did. All these observations are consistent with the moments lying along the chains (*b* axis).

The temperature dependence of the electrical resistivity of 4 different crystals is plotted in Fig. 3a. The reproducibility of the data from sample-to-sample is quite good. The resistivity was measured along the chains ($i//b$) and transverse to the chains ($i \perp b$) from 2-300 K. Measured along the chains, the resistivity is metallic with a noticeable downturn below the 140 K magnetic transition. The $i//b$ resistance at 40 mK was nearly the same as it was at 4 K, suggesting no further phase transitions (including superconductivity) down to that temperature. In contrast, the $i \perp b$ results are considerably less metallic with a broad hump near 140 K, constituting a remarkably anisotropic feature. No thermal hysteresis was observed across the transition, suggesting that the transition is continuous rather than first order.

The inset of Fig. 3a shows transverse MR data obtained with current applied parallel to the chains in magnetic fields to ±90 kOe. The MR ratio was calculated using the formula $MR = \rho/\rho_0 - 1$, where $\rho_0$ stands for the zero-field resistivity. Above the magnetic transition temperature, $NaV_2O_4$ shows a small degree of negative MR (<0.01) that is consistent with normal metallic behavior. Below the magnetic transition, the sign of the MR changes and its magnitude increases, indicating that the energy dependence of the scattering or the electronic structure changes at the magnetic transition.

Figs. 3b and 3c show the specific heat and thermopower of $NaV_2O_4$, respectively. Both quantities clearly show an anomaly at the 140 K magnetic ordering temperature. A Sommerfeld coefficient $\gamma = 6.91(4)$ mJ mol V$^{-1}$ K$^{-2}$ and Debye temperature $\Theta_D = 498.0(8)$ K were inferred from a least squares fit to $C_p/T$ vs. $T^2$ as shown in the inset to Fig. 3b. The experimental $\gamma$ is comparable to the calculated result (~4mJ mol V$^{-1}$ K$^{-2}$ for the FM solution and ~10mJ mol V$^{-1}$ K$^{-2}$ for the non-magnetic solution, see below), indicating that heavy quasiparticles are not being formed in NaV$_2$O$_4$ unlike the related compound LiV$_2$O$_4$ [12]. This is not surprising, as the magnetic entropy is eliminated via magnetic ordering rather than being transferred to the Fermi liquid. Below ~1 K, a strong upturn of specific heat was observed (not shown), which was most likely a nuclear Schottky contribution, hyperfine enhanced due to the magnetic ordering at higher temperature. Quantitative analysis of the magnetic entropy was accomplished by first subtracting the lattice specific heat of CaSc$_2$O$_4$ [13], an isostructural, non-magnetic analogue, from the zero-field specific heat of NaV$_2$O$_4$. After subtraction, the integrated entropy associated with the magnetic transition was ~6 J mol$^{-1}$ K$^{-1}$; this corresponds to only 40% of $R(\ln3+\ln2)$, expected if whole spins of V$^{3+}$ ($3d^2$: $t_{2g}^2$ $e_g^0$, $S$=1) and V$^{4+}$ ($3d^1$: $t_{2g}^1$ $e_g^0$, $S$=1/2) are localized and ordered at low temperature. It is likely that much of the entropy is removed via short-range ordering above the magnetic transition temperature.

The electronic structure of NaV$_2$O$_4$ was studied using the local spin density approximation of [14] density functional theory [15]. We used the WIEN2k package [16], which is based on the full-potential augmented-plane-wave method. Experimental lattice parameters and atomic coordinates were used for the calculations. The atomic radii were chosen as 2.1, 1.9, and 1.7 a.u. for Na, V, and O, respectively. The cut-off wave-number $K$ for wave functions in interstitial region was set to $RK$ = 7.5, where $R$ is the smallest atomic sphere radius. The integration over Brillouin zone was performed by a tetrahedron method with 132 k-points in the irreducible Brillouin zone. The non-magnetic density of states (DOS) is plotted in Fig. 4a. Note that V $d$-bands (-0.5 to 4 eV) are sharply separated into $e_g$-bands (1.8 to 4 eV) and $t_{2g}$-bands (-0.5 to 1.5 eV), reflecting the small degree of the VO$_6$ distortion. This result suggests that the $t_{2g}$ orbitals determine the low energy properties of NaV$_2$O$_4$. Moreover, because the $t_{2g}$ electrons remain highly itinerant along the chain direction, neither a spin-Peierls transition nor a spin density wave are likely to explain the magnetic transition at ~140 K.

The calculated Fermi energy lies near a major peak in the DOS, which suggests a magnetic instability. Indeed, we found a stable FM DOS solution (shown in Fig. 4b), which gained ~0.15eV per V atom from the non-magnetic total energy. The exchange splitting lifts the minority $t_{2g}$-bands above the Fermi energy and empties the bands completely, resulting in a half-metallic (HM) state. The estimated magnetic moment is 3.0 $\mu_B$ per primitive unit cell owing to the integer filling of the minority bands, suggesting the whole $t_{2g}$ electrons, $d^1+d^2$, fully contribute to the magnetism. The estimation is likely consistent with the result from the susceptibility study, however further studies of a role of spin fluctuations is needed to clarify nature of the magnetism. In addition, an antiparallel order state was tested as well; however, a stable DOS solution was not obtained. These results suggest strong instability toward a FM ordering in the present system, being consistent with the metallic transport property observed at low temperature. However, in contrast to the band structure calculation, bulk ferromagnetism was not experimentally observed. The facts thus suggest that the interchain magnetic coupling is rather antiferromagnetic while the intrachain coupling is ferromagnetic, preventing the appearance of net magnetic moment.

In strong contrast to the Stoner instability picture, it may be that double-exchange interactions of the $t_{2g}$ electrons are able to account for the experimental results; such a scenario has been discussed in the context of orthorhombic perovskite oxides [9,10,17]. In order to determine which of these models provides a better theoretical description, it is necessary to determine the level splitting between the lowest energy fully occupied orbital and the higher energy partially filled orbitals. Such studies require better treatment of on-site strong correlations using, for example, dynamical mean field method [18].

The most realistic magnetic model of NaV$_2$O$_4$ that accounts for the results presented here is that the integer magnetic moments order ferromagnetically along the chains, with the chains coupled antiferromagnetically. The spatial distribution of the spin current in that case would be a 1D analog of what has been predicted by a first principals calculation in the layered compound Na$_x$CoO$_2$ ($x \geq 0.75$) in a condition, in which HM layers are coupled antiferromagnetically [19-21].

In the HM chain model, the resistivity kink at ~140 K would be due to a Fisher-Langer singularity [22]. Transport perpendicular to the chains is hindered because there are no available states for the electrons to hop into, and thus the anisotropy and the non-metallic character of the resistivity observed in the $i \perp b$ measurements can be readily understood.

In summary, we have synthesized and characterized the basic physical properties of a new mixed-valent vanadium oxide, $NaV_2O_4$, which shows remarkable magnetic and electrical properties. The degree of the 1D anisotropy ($\rho_\perp/\rho_{//}$) is > 20 at low temperature, which is large for a transition metal oxide (c.f. ~6 in the double rutile-chains compound $BaRu_6O_{12}$ [23]). First-principles calculations indicate that HM ferromagnetism is found on the chains, whereas the behavior of the magnetization is consistent with AF order, with moments collinear with the chains.

We would like to thank Prof. W.E. Pickett (UC DAVIS) for helpful discussions and Dr. Y. Michiue (NIMS) for the X-ray data collection. The research was supported in part by the Superconducting Materials Research Project from MEXT, Japan, the Grants-in-Aid for Scientific Research from JSPS (18655080), the Futaba Electronics Memorial Foundation, the Murata Science Foundation, and Division of Materials Sciences and Engineering, Office of Basic Energy Sciences, U.S. Department of Energy, under contract DE-AC05-00OR22725 with Oak Ridge National Laboratory, managed and operated by UT-Battelle, LLC. Work at Los Alamos National Laboratory was performed under the auspices of the U.S. Department of Energy.

Table 1  Atomic coordinates and anisotropic displacement parameters of NaV$_2$O$_4$ at 293(2) K.

---

| Atom | $x$ | $z$ | $100U_{11}$ | $100U_{22}$ | $100U_{33}$ | $100U_{13}$ |
|------|-----|-----|-------------|-------------|-------------|-------------|
| Na | 0.24271(13) | 0.34620(10) | 1.28(6) | 1.35(6) | 1.22(6) | 0.02(4) |
| V1 | 0.08309(5) | 0.60374(4) | 0.94(2) | 0.81(3) | 0.58(3) | 0.013(15) |
| V2 | 0.06330(5) | 0.11198(4) | 0.93(2) | 0.86(3) | 0.59(3) | -0.013(15) |
| O1 | 0.2910(2) | 0.64828(16) | 1.09(9) | 0.97(9) | 0.90(9) | 0.05(6) |
| O2 | 0.3872(2) | -0.02053(17) | 0.97(9) | 0.97(9) | 0.74(9) | -0.05(7) |
| O3 | 0.3872(2) | 0.21757(18) | 1.09(9) | 1.05(9) | 0.74(8) | 0.04(7) |
| O4 | 0.0785(2) | -0.07120(18) | 1.08(9) | 0.91(9) | 0.67(8) | 0.03(6) |

---

Space group: *Pnma*, $a$ = 9.1304(11) Å, $b$ = 2.8844(4) Å, $c$ = 10.6284(13) Å, $z$ = 4, Cell volume = 279.91(6) Å$^3$.   The parameters $y$, $U_{12}$, $U_{23}$ are 0.25, 0, 0, respectively, for every atom.

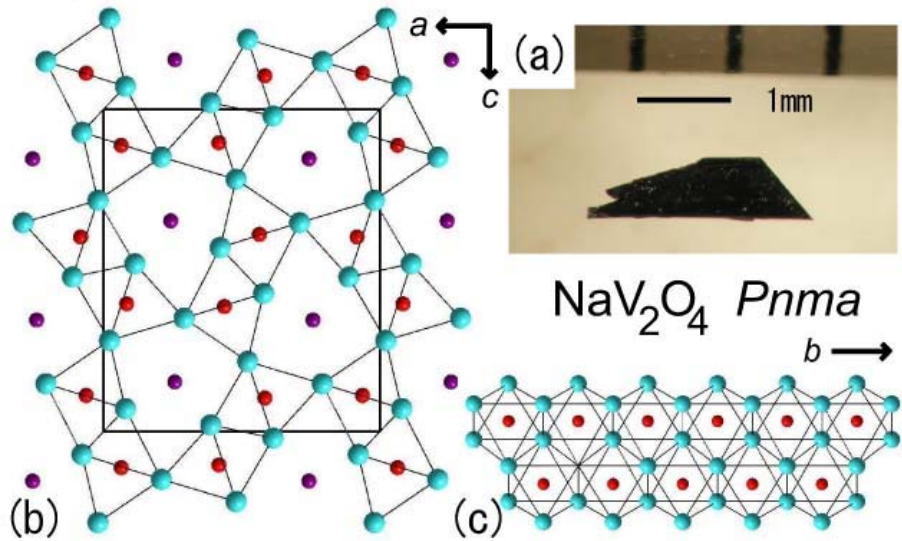

Fig. 1 (a) The crystal of NaV$_2$O$_4$ (1.779 mg) used for magnetization measurements. This is the largest crystal grown to date. (b) Schematic structural view of NaV$_2$O$_4$, drawn from the X-ray result, and (c) a local structural view of the double chain. The thick lines indicate the conventional orthorhombic unit cell, and the red, blue, purple circles indicate V, O, Na atoms, respectively.

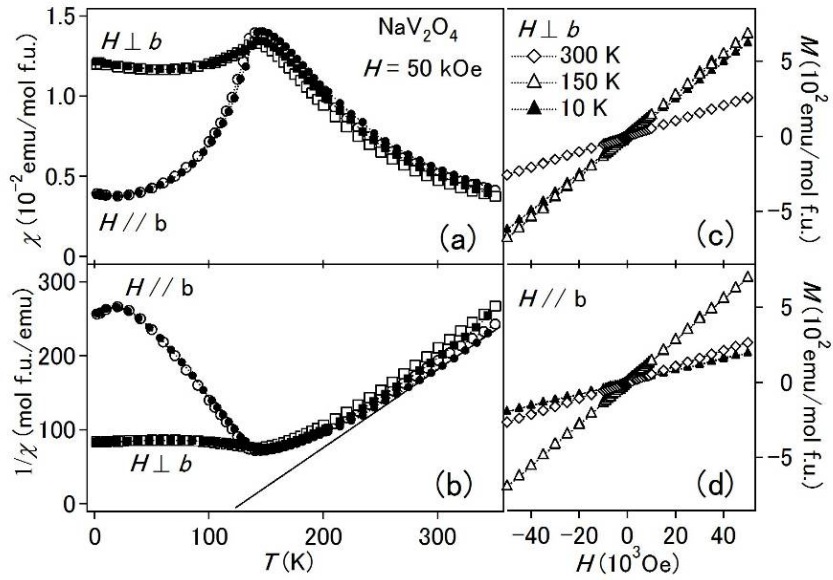

Fig. 2 (a,b) Temperature dependence of the magnetic susceptibility $\chi$ and $1/\chi$, and (c,d) isothermal magnetization of a crystal of $NaV_2O_4$ (see Fig. 1a). Open and closed symbols in the susceptibility curves represent zero-field cooling and field cooling data, respectively.

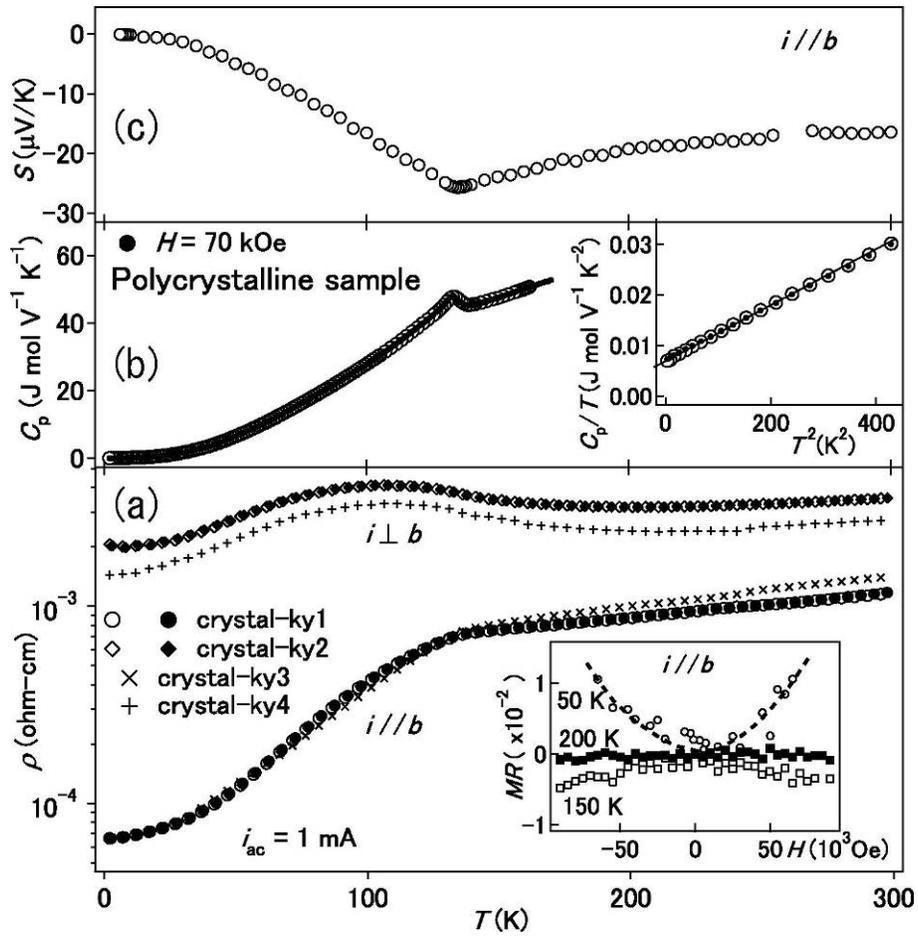

Fig. 3 (a) Temperature dependence of electrical resistivity, (b) specific heat, and (c) thermopower of $NaV_2O_4$. (Inset, a) MR ratios of $\rho_{//}$ at $H{\perp}b$, and (inset, b) a $C_p/T$ vs. $T^2$ fit of the low temperature specific heat data used to extract the Sommerfeld coefficient and Debye temperature.

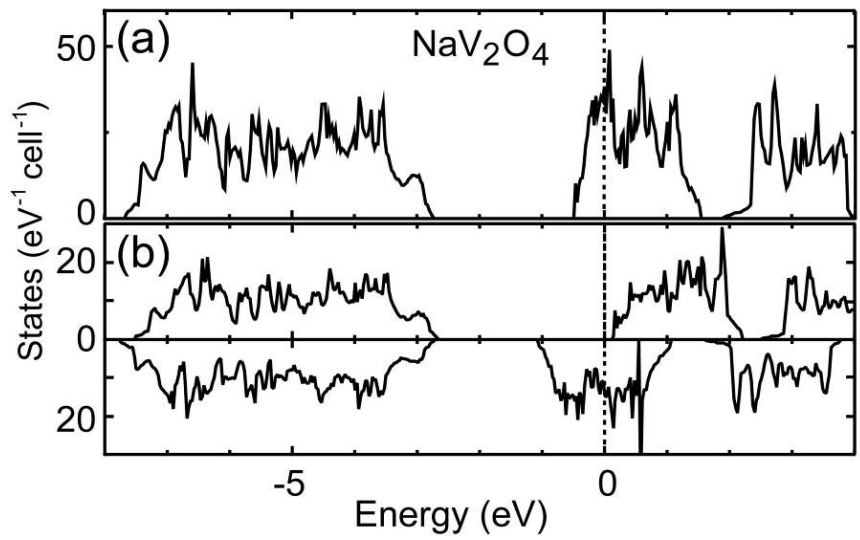

Fig. 4 (a) Non-magnetic DOS of NaV$_2$O$_4$. The sharp peak near $E_\text{F}$ suggests a magnetic instability. (b) The ferromagnetic DOS shows half-metallic ferromagnetism.